\documentclass[12pt,a4paper]{article}

\usepackage{amsmath}
\usepackage{amsfonts}
\usepackage{amssymb}

\begin{document}

\title{{\bf{\Large Covariant Anomalies, Horizons and Hawking
Radiation}}}
\author{
 {\bf {\normalsize Rabin Banerjee} \thanks{rabin@bose.res.in}}\\
{\normalsize S.~N.~Bose National Centre for Basic Sciences,}
\\{\normalsize Block JD, Sector III, Salt Lake, Kolkata -- 700098, India}
\\
\\}

\maketitle
\begin{center}

{\it This essay was awarded an Honorable Mention in the\\2008
Gravity Research Foundation essay competition.}

\end{center}

\begin{center}

\end{center}

\begin{center}
{\bf Abstract:}
\end{center}

Hawking radiation is obtained from anomalies resulting from a
breaking of diffeomorphism symmetry near the event horizon of a
black hole. Such anomalies, manifested as a nonconservation of the
energy momentum tensor, occur in two different forms -- covariant
and consistent. The crucial role of covariant anomalies near the
horizon is revealed since this is the {\it only} input required to
obtain the Hawking flux, thereby highlighting the universality of
this effect. A brief description to apply this method to obtain
thermodynamic entities like entropy or temperature is provided.

\pagebreak

\section{Introduction}

On quantising matter fields in a background spacetime containing an
event horizon -- for instance a black hole -- Hawking radiation is
obtained, a result that is classically unattainable. Ever since
Hawking's initial paper \cite{1}, there have been several
derivations \cite{2,3,4}, each with its own advantages or
disadvantages. While such derivations seem to reinforce Hawking's
original conclusion, none is completely general or truly clinching.

In this essay I will discuss a new approach to the Hawking effect
that is based solely on properties near the event horizon. The
method is essentially linked to the existence of gauge and
gravitational (diffeomorphism) anomalies near the event horizon, as
exemplified in recent approaches \cite{5,6,7,8}.

An anomaly signifies the breakdown of some classical symmetry due to
the process of quantization. The ubiquitous role of anomalies in
explaining various physical phenomena is well known \cite{9} and its
connection with the Hawking effect dates back to a seminal paper by
Christensen and Fulling \cite{3}. However, central role ascribed to
conformal symmetry is somewhat conceptually unpleasant since general
relativity is not conformally symmetric. Rather, the crucial
symmetry in general relativity is the invariance under general
coordinate transformations (diffeomorphism symmetry).

A couple of years back Wilczek and collaborators \cite{5,6} advanced
a new approach to the Hawking flux where diffeomorphism symmetry
plays a significant factor. They observe that, near the horizon,
black hole dynamics is effectively described by a two dimensional
chiral theory that breaks diffeomorphism symmetry. Requiring that
the complete theory with contributions near the horizon, outside the
horizon and inside the horizon be anomaly free, a condition is
obtained from which the Hawking flux is identified.

To put the above considerations in a proper perspective, it is
important to realise that there are two types of chiral anomalies --
covariant and consistent. Those satisfying the Wess Zumino
consistency condition are called consistent while those transforming
covariantly under the appropriate symmetry transformation are called
covariant. The covariant and consistent structures are different and
related by local counterterms \cite{9}. In fact this mismatch between the
covariant and consistent currents (or e.m. tensors) is the germ of
the anomaly. For an anomaly free theory the covariant and consistent
expressions for the currents (or e.m. tensors) are identical.

\section{Covariant anomaly and Hawking flux}

The anomaly method \cite{5,6} raises several issues, both technically
and conceptually. The flux is obtained there from the consistent
expression but the boundary condition involves the covariant form.
Note that the flux is measured at infinity where there is no
anomaly, so that covariant and consistent structures are identical.
Hence if the anomaly method is viable the flux should equally well
be obtainable from the covariant expression. Apart from being
conceptually clean, since only covariant expressions are involved,
it also entails considerable technical simplification. Shifts
between covariant and consistent expressions through counterterms,
as is mandatory in \cite{5,6}, are avoided. How this is done is now
illustrated following Banerjee and Kulkarni \cite{7,8}.

Consider a metric
$ds^2=f(r)dt^2-\frac{1}{f(r)}dr^2-r^2d\Omega^2_{(d-2)}$, where
$d\Omega^2_{(d-2)}$ is the line element on $S^{(d-2)}$, which
admits an event horizon at $f(r_h)=0$. As explained, the effective
theory near the horizon is chiral and two dimensional with
the metric $ds^2=f(r)dt^2-\frac{1}{f(r)}dr^2$. Here the covariant
diffeomorphism anomaly is given by $\nabla_\mu
T^\mu_{\nu(H)}=\frac{1}{96\pi}\frac{\varepsilon_{\nu\rho}}{\sqrt{-g}}\nabla^\rho
R$, where $T^\mu_{\nu(H)}$ is the covariant e.m. tensor near the
horizon, $\varepsilon_{\nu\rho}$ is the numerical antisymmetric
tensor and R is the Ricci scalar. Moreover $\sqrt{-g}=1$. In a
static background the anomaly is time like and one finds, for
$\nu=t$ \cite{7},
\begin{equation}\label{Y}
    \nabla_\mu T^\mu_{t(H)}=\partial_r T^r_{t(H)}=\partial_r
    A^r_t\,;\quad A^r_t=\frac{1}{96\pi}[ff''-\frac{1}{2}f'^2]
\end{equation}
This yields the solution,
\begin{equation}\label{X}
    T^r_{t(H)} = C+\int^r_{r_h}dr \partial_r A^r_t =
    A^r_t(r)-A^r_t(r_h).
\end{equation}
The integration constant $C$ is set to zero by imposing the boundary
condition that the e.m. tensor vanishes at the horizon
$T^r_{t(H)}(r_h)=0$. Such a condition is equivalent to the
regularity condition used in obtaining Hawking fluxes in Unruh
vacuum \cite{8}.

Initially ignoring the ingoing modes, the e.m. tensor is now written as a sum of two regions, near and outside the horizon. Taking the divergence of this tensor and using (\ref{Y}) naturally yields a Wess Zumino term that is interpreted as a contribution from the classically ignored ingoing modes. The redefined e.m. tensor, after including this contribution, is anomaly free provided $T^r_{t(0)}-T^r_{t(H)}+A^r_t=0$. Using (\ref{Y}) and
(\ref{X}) the Hawking energy momentum flux is obtained,
\begin{equation}\label{Z}
    T^r_{t(0)} = -A^r_t(r_h)=\frac{1}{192\pi}f'^2(r_h)=\frac{\pi}{12\beta ^2}
\end{equation}
where the final result is expressed in terms of the inverse Hawking
temperature $\beta$ by using
$k=\frac{2\pi}{\beta}=\frac{f'(r_h)}{2}$. This is just the desired
flux from blackbody radiation.

\section{New approach -- universality of Hawking radiation}

As seen above the reformulation of the anomaly method \cite{5,6} in
terms of covariant expressions only \cite{7,8} is simple and
straightforward, ironing out technical complexities and conceptual
issues. However certain problems still persist. The universality of
Hawking radiation requires that the flux be determinable only from
information at the horizon. The point is that, apart from the
anomalous Ward identity at the horizon, the normal Ward identity
outside the horizon is also required. Furthermore, it is also
necessary to interpret an additional Wess-Zumino term as a
contribution from the (classically irrelevant) ingoing modes. The
question is whether it is possible to derive the flux just from the
information of the chiral anomaly at the horizon. This is indeed so.

On observing the structure of the anomaly in (\ref{Y}) and imposing
asymptotic flatness of the metric $(\, f(r\rightarrow\infty)=1,
f'(\infty)=f''(\infty)=\ldots=0)$ it is found that $A^r_t$ vanishes
in the $r\rightarrow\infty$ limit. Hence the anomaly also vanishes
in this limit. This implies that the $r\rightarrow\infty$ limit of
$T^r_{t(H)}$ would correspond to the anomaly free expression
associated with $T^r_{t(0)}$. Consequently the Hawking flux, which
is measured at infinity, will be given by,
\begin{equation*}
    T^r_{t(0)}=T^r_{t(H)}(r\rightarrow\infty)=-A^r_t(r_h)=\frac{\pi}{12
    \beta ^2}
\end{equation*}
which follows from (\ref{X}) on using $A^r_t(r\rightarrow\infty)=0$.
This reproduces (\ref{Z}).

The Hawking flux is therefore obtained solely from information at
the horizon with the minimum of effort. Inclusion of gauge fields,
as required in the case of charged black holes, poses no problems.
Also, higher spin fluxes can be computed in an identical way leading
to the full Hawking spectrum.

An important open issue in this context is the computation of
thermodynamic entities like entropy or temperature of a black hole.
Arguments will be given to show that this should be feasible. The
present method has similarities with conformal field theory
techniques \cite{10} used to study black hole thermodynamics. There the
entropy is computed by the Cardy formula \cite{11} which requires both the
central charge and the conserved charge. Something analogous can be
done here. In the conformal field theory approach, the central
charge (or conformal anomaly) is obtained from the central extension
of the Virasoro algebra by redefining the usual diffeomorphism
generators by `stretched horizon' constraints \cite{12}. Likewise,
the conserved charge is obtained from a boundary term needed to make
the redefined generators differentiable. A central extension is
naturally obtained in the present analysis without the need of any
redefinitions, due to the chiral anomaly near the horizon. This has
already been computed in the literature \cite{9}, though in a
different context. Also, the conserved charges, which are roughly
analogous to energies, can be computable, as outlined in \cite{8}
from two dimensional effective actions subjected to the boundary
conditions used here.

It appears that different manifestations of the anomaly provide a
unifying picture that illuminates the universality of the Hawking
effect.


\begin{thebibliography}{99}

\bibitem{1}
S. Hawking, Comm. Math. Phys. {\bf 43} (1975) 199

\bibitem{2}
G. Gibbons and S. Hawking, Phys. Rev. {\bf D15} (1977) 2752

\bibitem{3}
S. Christensen and S. Fulling, Phys. Rev. {\bf D15} (1977) 2088

\bibitem{4}
M. Parikh and F. Wilczek, Phys. Rev. Lett. {\bf 85} (2000) 5042

\bibitem{5}
S. P. Robinson and F. Wilczek, Phys. Rev. Lett. {\bf 95} (2005) 011303

\bibitem{6}
S. Iso, H. Umetsu and F. Wilczek, Phys. Rev. Lett. {\bf 96} (2006) 151302

\bibitem{7}
R. Banerjee and S. Kulkarni, Phys. Rev. {\bf D77} (2008) 024018

\bibitem{8}
R. Banerjee and S. Kulkarni, Phys. Lett. {\bf B659} (2008) 827

\bibitem{9}
R. Bertlmann, Anomalies in Quantum Field Theory, Oxford Sciences, Oxford, 2000;
K. Fujikawa and H. Suzuki, Path Integrals and Quantum Anomalies, Oxford Sciences, Oxford, 2004.

\bibitem{10}
For a review, see S. Carlip, Gen. Rel. Grav. {\bf 39} (2007) 1519

\bibitem{11}
J. A. Cardy, Nucl. Phys. {\bf B270} (1986) 186;
H. W. J. Bl\"ote, J. A. Cardy and M. P. Nightingale, Phys. Rev. Lett. {\bf 56} (1986) 742

\bibitem{12}
S. Carlip, Int. J. Theor. Phys. {\bf 46} (2007) 2192

\end{thebibliography}
\end{document}